\documentclass{elsart}
\usepackage{amsmath}
\usepackage{amssymb}
\usepackage{graphicx}
\usepackage{bm}% bold math

\begin{document}

\begin{frontmatter}
\title{Thermodynamics of volume collapse transitions in cerium and related compounds}

\author{S. Bustingorry, E. A. Jagla}
\address{Consejo Nacional de
Investigaciones Cient\'{\i}ficas y T\'ecnicas\\ Centro At\'omico Bariloche,
8400 Bariloche, Argentina}
\author{J. Lorenzana}
\address{SMC, INFM, ISC, CNR, Dipartimento di Fisica,  
 Universit\`a di Roma ``La Sapienza''
 P. Aldo Moro 2, 00185 Roma, Italy}
\begin{abstract}
We present a non-linear elastic
model of a coherent transition with discontinuous volume change in an isotropic
solid.  
The model reproduces the anomalous thermodynamics typical of
coherent equilibrium including intrinsic hysteresis (for a pressure
driven experiment) and a negative bulk modulus. The novelty of the
model is that the statistical mechanics solution can be easily 
worked out. We find that coherency leads to an
infinite-range density--density interaction, which drives
classical critical behavior. The pressure width of
the hysteresis loop shrinks with 
increasing temperature, ending at a critical point at a temperature related 
to the shear modulus. The bulk modulus softens with a 1/2 exponent at
the transition even far from the critical point.  Many well known features of 
the phase diagram of Ce and related systems are explained by the model.
\end{abstract}

\begin{keyword}
% keywords here, in the form: keyword \sep keyword
%\pacs{71.28.+d}{Narrow-band systems; intermediate-valence solids}
%\pacs{64.70.Kb}{Solid-solid transitions}
%\pacs{68.35.Rh}{Phase transitions and critical phenomena (solid surfaces/interfaces)}
%\pacs{71.10.Hf}{Non-Fermi-liquid ground states, electron 
%phase diagrams and phase transitions in model condensed matter systems}
volume collapse \sep mixed valence systems \sep nucleation \sep Ce \sep SmS 
\PACS 71.28.+d \sep 64.70.Kb \sep 68.35.Rh \sep 71.10.Hf
\end{keyword}
\end{frontmatter}

\section{Introduction}

Room temperature $\gamma$ cerium shows a well known transition to 
isostructural $\alpha$ cerium at %0.8GPa with a 18\% (Franceschi)
$\sim$ 0.8GPa  with a 17\% reduction of the volume.
 Similar (though not necessarily isostructural) transitions occur in 
other $f$-electron systems usually associated with mixed-valence 
behavior\cite{gsc79,law81}. 
Volume-collapse transitions  were also predicted to occur in 
colloidal systems\cite{bol94} and close to 
first order electronic transitions when mesoscopic inhomogeneous electronic 
states are frustrated by the
long-range Coulomb interaction\cite{lor01I,lor02}.
In Ce and SmS the volume collapse transition ends at a critical point,
 making of these and related systems the solid state analog of a van der
Waals liquid-gas system.

The thermodynamics of coherent structural transitions have been shown to 
be highly anomalous\cite{cah61,cah84,roi84,roi85,fra99,wag74,bsch95,roi78,roy97,roy99,kha83}.
In the mixed phase the system behaves non extensively
and violates usual stability criteria for extensive systems\cite{cah84,roy97,roy99}. In
particular, it possesses negative compressibility\cite{roy97,roy99}. For an experiment 
in which the pressure acts as the control variable 
usual nucleation is forbidden within a finite pressure window.
This pressure window implies an intrinsic hysteresis loop 
as pointed out by Roytburd\cite{roi84,roi85,roy97,roy99,roy97phb} 
and discussed also in metal--hydrogen systems\cite{roi85,bsch95}.
Outside the pressure window the transition can 
occur, but only irreversibly with an intrinsic jump of the
thermodynamic potential. 

In this work we present a simple non-linear elastic model 
describing a coherent volume-collapse transition 
in an isotropic solid. The assumed elastic coherence
leads to a distance-independent density--density interaction 
analogous to compositional effects in alloys\cite{cah61,cah84,roi84,roi85,fra99} 
and hydrogen density--density interaction in metals\cite{wag74}.
Mean-field (MF) theory becomes exact and the statistical mechanics
solution of the model can be easily worked out, which constitutes the
main novelty of the model. The solution reproduces the 
above mentioned anomalous thermodynamics. In addition some novel results are found. 
The extent of the pressure hysteresis reduces as a function of temperature,
vanishing at a critical point. The temperature of the critical 
point is directly related to the  shear modulus of
 the material. 
We show that critical exponents are MF like, 
and this remains true  even in the presence of short-range interactions, providing a
straightforward explanation to old experiments in 
mixed-valence systems\cite{law81}. 
We discuss under what conditions
the transition occurs with a softening of the bulk modulus and discuss
its critical behavior.

\section{Model}

In the present model, the order parameter (OP) 
is taken to be the dilation strain         
$e\equiv\sum_{i=1}^d \epsilon_{ii}$
 where $\epsilon_{ij}$ are the components of the infinitesimal
strain tensor\cite{cha95}
 $\epsilon$ and $d$ is the 
dimensionality of the system.
The dilation strain  describes
changes in the local specific volume with respect to a reference
state, and it is defined in such a way that 
if $V_0$ is the volume of an original piece of material (at $e=0$),
 then the volume for $e\ne 0$ is $V=V_0(1+e)$.

Bistability in the volume is described by a 
coarse-grained energy functional $H_1$.
If we assume that upon collapse, the system passes from a specific
volume $v^+_0$ to some lower value $v^-_0$, then 
an appropriate description of the transition
will contain a local term of the form 
\begin{equation}
\label{Eq:fgl}
H_{1}=\int d^d{x} \left( -\frac{b}2 e^2 
+ \frac{c}4 e^4 + \frac{\kappa}2 |\nabla e|^2 -ae\right),
\end{equation}
Here the reference state (for which $e=0$) was taken to correspond
to a specific volume $v=(v^-_0+v^+_0)/2$, and the relative volume change 
at the transition (at zero temperature) is given by
\begin{equation}
\frac{\Delta v}{v}(T=0)\equiv 2\frac{v^+_0 -v^-_0}{v^+_0 +v^-_0}=2\sqrt{\frac
  bc}.
\label{dvv}
\end{equation}
The linear term in (\ref{Eq:fgl}) is included in order to have the
 transition occurring at a finite pressure.
$H_1$ alone describes an incoherent volume collapse transition with nominal
volume change $\Delta v/{v}$ and at a nominal pressure $P=a$. 
Coherence effects will change this simple picture. 

In a coherent transition non-uniform dilation strains are accompanied by shear
strains. The latter give an additional elastic 
contribution $H_{h}$ to the total energy. 
To lowest order we take this 
contribution to be harmonic, and for the assumed isotropic case 
we have: 
\begin{equation}
  \label{Eq:fh}
H_h= \mu \int d^d{x} \sum_{i,j=1}^{d}\widetilde{\epsilon}_{ij}\,^2.
\end{equation}
where $\mu$ is the shear modulus and 
 the traceless strain  tensor $\widetilde{\epsilon}_{ij}\equiv\epsilon_{ij}
-\delta_{ij}e/d$ has been defined.

%For a given configuration of the order parameter $e({\bf x})$ one can find the
% configuration of shear strains $\widetilde{\epsilon}_{ij}({\bf x})$
%that minimize the energy. In this way, an energy functional dependent on $e({\bf x})$
%alone will be obtained. 
%The shear strains are said to ``accommodate'' to the order parameter so 
%they will be also referred to as the ``accommodation strains''.

The coefficients in (\ref{Eq:fgl}) and (\ref{Eq:fh}) are assumed to be
temperature and pressure independent, i.e., 
Eq. (\ref{Eq:fgl}) has to be considered as an energy functional (rather than a Ginzburg--Landau 
free energy). The phase transition will be derived from the present energy
functional by its 
introduction into a partition function and a full statistical mechanics calculation. 
Note also that we are assuming that the two phases have the same bulk and shear modulus.

\begin{figure}
\includegraphics[width=7cm,trim=0 150 0 50,clip=true]{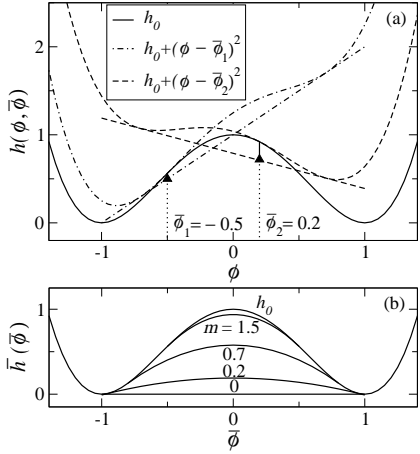}
\caption{(a) $\bar\phi$-dependent common tangent construction. The cases 
$\bar\phi=-0.5$ and $0.2$ are illustrated ($m=1$). The triangles indicate the
minimum value of $h$ attainable for the corresponding values of $\bar \phi$. 
(b) Analytical results for $\bar h(\bar\phi)$ as a function 
of $\bar\phi$ for $\gamma=0$ and
different values of $m$, as indicated. For $m>2$ the solution is 
$\bar h(\bar\phi) = h_0(\bar\phi)$.}
\label{fig:fdf}
\end{figure}

\section{Zero temperature theory }
In a coherent structural phase transition the
strain tensor must fulfill the St. Venant compatibility
constraints
$
\nabla \times (\nabla \times {\bf \epsilon})^{\dagger}=0
$
\cite{roi84,roi85,roi78,roy97,roy99,roy97phb,cha94,kar95,ras01,kle02}.
This leads to constrains between the shear strains and the
OP. We take periodic boundary conditions and work with the
Fourier transform of the strain tensor, $\epsilon({\bf k})$.

The ${\bf k}=0$ and  ${\bf k}\ne 0$ components have to be
treated separately\cite{cho96}. Indeed a uniform strain satisfies the
St. Venant compatibility constraints  trivially and therefore
it is unconstrained. For ${\bf k}\ne{\bf 0}$ the shear strains
can be eliminated in favor of $e$ by minimizing
$H_h$ with respect to the shear strains as shown in the Appendix.
The shear strains are said to ``accommodate'' to the order parameter so
they will be also referred to as the ``accommodation strains''.
As shown in the Appendix this allows us to express Eq.~(\ref{Eq:fh})
as a function of the OP alone:
\begin{equation}
  \label{Eq:fhde1k}
H_h=\left(1-\frac1d\right)\mu\frac1{L^d}\sum_{{\bf k}\ne{\bf 0}} e({\bf k})e(-{\bf k}).
\end{equation}
Here $L$ is the linear dimension of the system and  $e({\bf k})$ are
the Fourier components of $e$.
A crucial point is that, since the ${\bf k}={\bf 0}$ term is unconstrained, it is
excluded from the sum.
We can formally extend the sum to all 
${\bf k}$ by explicitly subtracting the missing term. 
% For ${\bf k}\ne{\bf 0}$, the Fourier components of the
One obtains in real space:
\begin{equation}
  \label{Eq:fhde1}
H_h=\left(1-\frac1d\right)\mu \int d^d{x} [e({\bf x})-\bar e]^2 %\\
\end{equation}
with $\bar e\equiv\int d^d{x} e /L^d$.
A similar result has been obtained in $d=2$ 
in the context of the dislocation mediated melting 
problem\cite{cho96}.

Eq.~(\ref{Eq:fhde1}) manifestly satisfies the Bitter--Crum theorem of 
coherent mixtures which states that the elastic energy is 
independent of the shape and location of the phases\cite{fra99}.

We define now $\phi\equiv e \sqrt{c/b}$, $m\equiv 4(d-1)\mu/(db)$ and 
$\gamma\equiv 2\kappa/b$. Up to an additive constant the coarse grained energy 
density $h$ [from Eqs.~(\ref{Eq:fgl}) and (\ref{Eq:fhde1})]
in units of the height of the double well barrier [$b^2/(4c)$] reads:
\begin{equation}
\label{Eq:gamma}
h(\phi,\bar \phi)= h_0(\phi)+\gamma |\nabla \phi|^2 + m (\phi-\bar \phi)^2
\end{equation}
with $h_0(\phi)\equiv (1-\phi^2)^2$ being the double well in the dimensionless
volume $\phi$. The linear term in the OP
has been eliminated by redefining the origin of pressure, i.e., the transition pressure is now
zero.
The parameter $\gamma$ 
%has units of inverse length squared and
fixes the width of the interface
between the low and high volume phases in a mixed state. 
The energy density 
Eq.~(\ref{Eq:gamma}) couples the values of $\phi$ at different 
spatial positions only through $\bar\phi$, namely it corresponds
effectively to a MF or an infinite range interaction.
We stress however, that this is the exact energy of the model, 
the mean-field contribution coming from the
harmonic elastic energy of the accommodation strains, Eq.~(\ref{Eq:fhde1}).

We first discuss the $\gamma = 0$ and zero temperature case.
Mixed phase solutions for a constrained total volume  
obey the familiar common tangent construction:
$\partial h(\phi,\bar \phi)/\partial \phi|_{\phi=\phi_-}=
\partial h(\phi,\bar \phi)/\partial \phi|_{\phi=\phi_+}=
[h(\phi_-,\bar \phi)-h(\phi_+,\bar \phi)]/(\phi_- -\phi_+)$
where  $\phi_{\pm}=\pm\sqrt{(1-m/2)}$ and the $+$ ($-$) corresponds 
to the OP of the expanded (collapsed) phase.
  The only difference with a conventional phase separation computation
is that the construction is $\bar \phi$-dependent because 
$h$ depends on the global variable $\bar \phi$ [see fig.~\ref{fig:fdf}(a)]. 
This dependence makes the total energy of the mixture 
non additive.  In fig.~\ref{fig:fdf}(b) we report the resulting
average energy density $\bar h(\bar \phi)$.

For $m>2$ the mixed solution does not exist and the system remains 
always uniform. For $m<4/3$ the mixed state solutions produce a shift
of the spinodal points of the system [fig.~\ref{fig:fdf}(b)] 
whereas for $m>4/3$ the spinodal
points coincide with the  spinodal points of $h_0$.

The dimensionless pressure $p$ can be calculated as 
$p=-\partial \bar h(\bar \phi)/\partial
\bar\phi$. The physical pressure $P$
is obtained from $p$ as $P=p \sqrt{b^3/(16c)} +a$. 
The nominal transition pressure
is $p=0$ (i.e., $P=a$). In the mixed phase the free energy has negative curvature
so the compressibility is negative [fig.~\ref{fig:fdf}(b)]. This is
also true for uniform phases for $m>4/3$ with volume between the 
spinodal points. 
%by $h(\bar\phi)= -m(\bar\phi^2-1+m/4)$ (see
%fig.~\ref{fig:fdf}(b)), the dimensionless pressure is $p=2 m \bar
%\phi$ and the compressibility is negative$(\propto -2m)$. 
This thermodynamic state is stable at constant volume with the 
pinned boundary conditions of Ref.~\cite{ber76,lak01}. We mention that 
this violation of usual stability criteria is common in other contexts
where long-range forces appear\cite{veg02}.

\section{Finite temperature theory }
For the model with $\gamma=0$, it is straightforward to obtain 
the exact partition function and 
the exact $\bar\phi-p$ dependence at finite temperatures.
Let us rewrite for convenience the energy density $h$ in the form
\begin{equation}
\label{Eq:mf}
h(\phi,\bar \phi)= h_0(\phi)+ m (\phi-\bar \phi)^2 +p\phi,
\end{equation}
where we set $\gamma =0$ and introduce explicitly the pressure term. 

To solve the model we have to consider the
minimum volume that can fluctuate independently in the system. 
This volume will be noted $\xi^d$, with $\xi$ an atomic length scale.
The energy is written in
discretized form as a sum over $N$ cells of volume 
$\xi^d$ with $L^d=N\xi^d$: 
$$H=\frac{b^2}{4c} \xi^d \sum_i^N h(\phi_i,\bar \phi).$$
The partition function can be written as $Z=z^N$ with
$$z(\bar\phi)=\int d\phi e^{-h(\phi,\bar \phi)/t},$$
where $t$ is a dimensionless temperature defined as: $t=k_BT 4c/(b^2\xi^d)$.
The value of $\bar \phi$ must be computed self-consistently as:
\begin{equation}
  \label{Eq:selfcons}
\bar \phi=\frac 1{z(\bar\phi)} \int d\phi \phi e^{-h(\phi,\bar \phi)/t}.
\end{equation}

\begin{figure}[bp]
\includegraphics[width=10cm,clip=true]{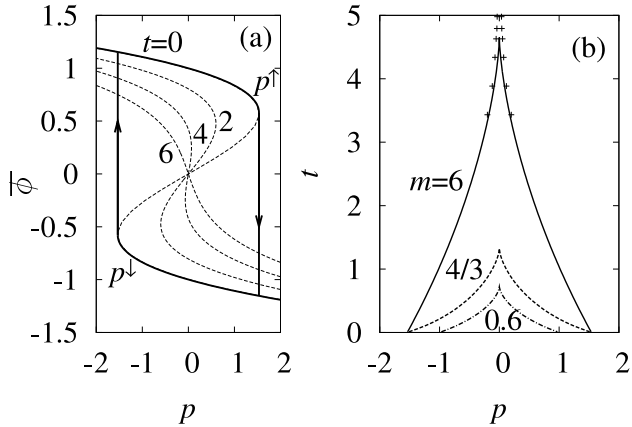}
\caption{(a) $\bar \phi$ vs. $p$ for $m=6$ and different values
of the dimensionless temperature $t$, as indicated. 
(b) $t$-$p$ phase diagram for $\gamma=0$ and 
different $m$ values, as indicated. The interior of the wedge
(for each value of $m$) is a region in which nucleation is forbidden.
The crosses are the experimental width of the hysteresis
loop for Sm$_{0.87}$Gd$_{0.13}$S~\protect\cite{ras78} rendered
dimensionless with theoretical estimates of $b$ and $c$.}
\label{fig:tem}
\end{figure}

In fig.~\ref{fig:tem}(a) we show
the isotherms in the $\bar \phi$-$p$ plane for $m=6$ for different 
values of the dimensionless temperature $t$. 
Controlling the total  volume the system would follow the dashed
 lines. 
For $t=0$ we also show the hysteresis loop (full line) 
that the system would follow in a pressure controlled experiment. 
 Starting from the high volume phase at low pressure and increasing 
 $p$ the system remains in this phase  until the spinodal pressure
 $p^{\uparrow}$, i.e. until the limit of stability of the phase where
 the  volume suddenly collapses. Reducing the pressure the system
 suddenly expands at $p^{\downarrow}$.
 A conventional (incoherent) first-order
 phase transition driven by pressure at a very low rate 
 (so that nucleation occurs) would follow 
 a vertical line at $p=0$.
 This is not possible here because the infinite range 
 interaction makes the critical drop radius\cite{cha95} diverge, and
 the nucleation energy-barrier scale as the                     %<<<<<<<<<<<<<<<<<
 volume of the system, a well known fact in structural transformations\cite{kha83}.   %<<<<<<<<<<<<<<<<<

The pressure width ($\Delta p$) and volume width ($\Delta \bar\phi$) 
of the hysteresis loop as
computed from Eq. (\ref{Eq:selfcons}) are reduced at finite temperatures 
up to the point where they vanish, which determines the
critical point coordinates $(p_c=0,t_c(m))$. The  pressure
width as a function of temperature determines a wedge of forbidden nucleation
[shown in fig.~\ref{fig:tem}(b) for different values of $m$] 
around the critical pressure $p_c=0$. 

Because of the mean-field character of the model the isotherms
are analytic functions when seen as pressure vs. volume plots
[rather than volume vs. pressure as in Fig.~\ref{fig:tem} (a)].
Therefore close to the spinodal, inverting the analytic behavior, 
one finds that the volume behaves as 
$[p-p^{\uparrow,\downarrow}]^{1/2}$. 
 The compressibility ($\propto \partial\bar \phi/\partial p$) diverges at  
the transition even far from the critical point with a 1/2 exponent.

The solution in the case of finite $\gamma$ and $m$ 
is given implicitly by  
$\bar\phi=\bar\phi_0(p+m\bar\phi)$ where 
$\bar\phi_0(p)$ is the equation of state of 
the same model with $m=0$. This corresponds to
the well studied $\phi^4$-model which is in the 
universality class of the Ising model\cite{cha95}. 
We can provide here only a description of the results that are
obtained, a full detailed solution will be presented elsewhere.
If $m>4/3$ the qualitative description of the forbidden nucleation 
region are completely equivalent to the $\gamma=0$ case, i.e., the previously discussed
pressure-temperature phase diagram and in particular the description of the forbidden 
nucleation wedge are not qualitatively affected.
For $m<4/3$ and sufficiently low temperatures, by driving the system through the
transition, the limit of forbidden nucleation is reached before the 
stability limit.
Therefore, there exists a pressure region, outside the window of forbidden
nucleation, where the system
can still remain in a metastable state 
until reaching the spinodal pressure, and
a standard nucleation and growth process
may occur due to the surface energy contribution of the gradient term. 
However, it is possible to show that this nucleation and growth region 
never reaches the critical point.
From this we can conclude that classical critical exponents are obtained also for $m<4/3$.
Thus we find that critical exponents take the mean-field values for both $\gamma=0$
and $\gamma>0$ . 
This is also in agreement with the $d=2$ theoretical analysis of 
Ref.~\cite{cho96}.

\section{Comparison with experiments}

A wedge where nucleation does not occur is 
observed in Ce\cite{gsc79,law81}, SmS\cite{ras78} and related
 systems. Interestingly it is also observed in a
volume-collapse transition in amorphous ice\cite{mis94}.
However without detailed studies it is difficult 
to be sure that it does not arise from the usual hysteresis
due to a finite driving rate, as occurs in other first order transitions. 
Other indicators, however, point more strongly to the applicability
of our theory.

In agreement with the present theory MF  critical exponents have been
observed in Ce systems\cite{law81}.
The connection with long-range strain effects
has long been suspected\cite{law81} 
but, to the best of our knowledge, has not received a microscopic 
explanation in mixed-valence systems.

In contrast with usual first-order phase transitions, where 
hysteresis is not intrinsic, one can define here a critical exponent for 
the pressure width of the hysteresis loop: $\Delta p\propto
|t-t_c|^{3/2}$. In particular the curvature of the boundaries 
close to the critical point must be
convex. This is in agreement with the
experimental data on Gd doped SmS\cite{ras78} shown in
Fig~\ref{fig:tem}(b) supporting our conclusion that  
$\Delta p$ is an intrinsic property of the system.

Also in contrast with usual first-order transitions, critical behavior 
(although classical like) can be easily observed  far from the critical point.
The divergence of the compressibility discussed above
implies that the bulk modulus vanishes as $|p-p^{\uparrow}|^{1/2}$ and 
jumps discontinuously to a finite value for $p>p^{\uparrow}$.
Remarkably, after this work was completed and a preliminary version of it was posted\cite{bus03} this
prediction was experimentally verified in Ce\cite{jeo04}. 

 Decreasing the pressure, the jump and the square root singularity change 
sides and occur at $p^{\downarrow}$.
Similar behavior is found as a function of temperature.
%In this case the boundaries of the forbidden nucleation 
%region coincide with the spinodal curves. 
% One can show
%that this is also true with a nonzero value of $\gamma$  
%therefore 
Also an anomaly 
measured in YbInCu$_4$ on cooling resembles our prediction when plotted in 
an appropriate scale (see inset of fig.~4 in Ref.~\cite{bkin94}). 
This also 
explains the appearance of precursor effects which have been very puzzling 
in the mixed-valence literature and have led some authors to interpret the 
transition as second  order\cite{bkin94}.

A divergent compressibility implies the unusual property
that the Poisson's ratio will become negative and approach $-1$.
 A negative Poisson's ratio 
is indeed observed in the SmS system and is generally associated 
with mixed-valence behavior\cite{sch98}. 
Since the unusual elastic properties we find (including negative
compressibility) occur in extensive temperature
and volume ranges which can additionally be tuned by alloying 
and external fields we conclude that mixed-valence systems are ideal 
candidates for applications like  induced hardening and giant damping
(c.f. Ref.~\cite{lak01}).

In mixed-valence systems the double well is attributed to the 
anomalous contribution of $f$-electrons to the crystal 
binding\cite{all82,joh95}.
For a  realistic description it may be necessary to take the parameters in
Eq.~(\ref{Eq:fgl}) to depend on temperature to take into account  
temperature effects of the  $f$-electrons. In fact a trivial linear 
and quadratic temperature dependence of $a$ fits our symmetric 
phase diagram to the asymmetric forms of Ce\cite{gsc79,law81} 
and SmS\cite{law81,ras78}. 
%Such a dependence can be related to the entropy 
%jump between the two faces in pseudoEquilibrium according to a 
%Clausius-Clayperon analysis.  For our symmetric double well such a jump 
%can have only electronic origin. 

The temperature dependence of $a$ does not affect
 the critical temperature, but $b$ does.
Indeed in current explanations of the critical  point in $f$-electron
systems this electronic effect is crucial whereas the shear 
rigidity does not play a relevant role\cite{all82,joh95}. 

To check the relevance of the temperature scale set by the shear modulus
in determining real values of $T_c$ we 
take the opposite point of view, i.e., we neglect the temperature 
dependence of the coefficients in  Eq.~(\ref{Eq:fgl}) and take into
account the effect of the shear rigidity alone. 

We can eliminate the
 unknown coefficients $a$ and $b$ in favor of the relative volume  
$\Delta v(T)/v$ and pressure jump  $\Delta P(T)$ at a given
 temperature $T$. $\Delta v(T)/v$  is defined at the pseudo
equilibrium pressure (practically the average between 
$P^\uparrow$ and    $P^\downarrow$).
Restoring dimensions [$k_B T=t \xi^d  b^2/(4c)$] the critical temperature
for $m\gg 4/3$ is given by:
\begin{equation}
  \label{Eq:tc}
T_c=T+\frac{\delta T+ \delta T'}2+
\sqrt{ \left[\frac{\delta T + \delta T'}2 \right]^2+  T \delta T' 
}  
\end{equation}
$k_B\delta T'\equiv C\Delta P(T)\frac{\Delta v(T)}{v}  \xi^d$,
$k_B \delta T\equiv\frac{d-1}{6d}  \xi^d \mu  \left( \frac{\Delta
    v(T)}{v}\right)^2 $ and $C\simeq 0.137$. 

This relation between $\mu$ and $T_c$ is rooted in the fact 
that the critical point along the critical isochore 
can be seen as an order--disorder transition governed by the 
rigidity of the lattice parametrized by $\mu$, in strong contrast
with current theories of the critical point in Ce\cite{all82,joh95}.
A similar relation between $\mu$ and $T_c$ has been obtained for 
metal hydrides\cite{wag74}. 

For the  $\gamma=0$ case, $\xi^d$ should be assimilated
to the atomic volume. In the case $\gamma>0$, this volume 
can be greater depending on the strength of the short distance interactions 
parametrized by $\gamma$. In any case we expect it to be of the 
order of a few formula unit volumes $v_f$.  
To make a rough estimate we took $\xi^d=2 v_f$  and we obtain 
good agreement with the experimental values (Table~\ref{tab:par}).
\begin{table}
\caption{Parameters at $T=293$K used to estimate $T_c$ [K]. 
$\Delta P$ and  $\mu$ are given in GPa. $\mu$ is an average value. 
Experimental data for Sm$_{1-x}$Gd$_x$S after Ref.~\cite{ras78} except
for $\mu$ which is taken at $x=0$\cite{hai84}. We caution that the experimental
values of $T_c$ are not well established and vary as much as 10\% for different authors. }
\label{tab:par}
\begin{center}
   \begin{tabular}{llllllll}
           & $\mu$ & $\Delta v/v$& $\Delta P$ & $v_f$ [\AA$^3$]&   $T_c^{\rm Exp.}$ & $T_c^{\rm Theory}$ &$m$ \\
Ce\cite{gsc79,law81}&13 & 0.17   &   0.2      & 31.3           &   550      &     530    & 5     \\
SmS& 40& 0.105& 0.2   & 53.3                       &   700      &     702    & 12    \\
Sm$_{0.87}$Gd$_{0.13}$S& & 0.075 &0.2 & 50.5                   &   543      &     508    & 6     \\ 
   \end{tabular} 
\end{center}
\end{table}

Since the relation between $\xi^d$ and $v_f$ goes beyond our model 
we cannot exclude  temperature effects on the coefficients of 
Eq.~(\ref{Eq:fgl}) contributing to $T_c$\cite{all82,joh95}; 
however, the fact that we obtain the right 
order of  magnitude with a  $\xi^d$  of the order of $v_f$ implies that the
scale set by the shear modulus can not be neglected in $T_c$. 

\section{Conclusions}
In conclusion we have presented a simple solvable model of an isostructural
volume collapse transition, the solid-state analog of the  van der Waals
 transition. The model reproduces the anomalous thermodynamics found
 in other approaches\cite{cah61,cah84,roi84,roi85,fra99,wag74,bsch95,roi78,roy97,roy99,kha83}.
In particular, the transition is intrinsically irreversible 
with a discontinuity of the thermodynamic potential. % making of it a  
The main novelty of our treatment is that the temperature is 
explicitly put in the problem (rather than indirectly through 
the dependence of the parameters of the model). As a consequence we can
discuss the critical behavior of the transition 
close to but also far from the critical point.
In particular we find a softening of the bulk modulus as the
transition is approached. 

In many respects the physics is similar to spinodal decompositions of
alloys; however, an important difference is that in
 the present case the order parameter is not conserved,
slow atomic diffusion is not involved, and
 the kinetics of the problem is very fast.  

The softening effects that we find in a transition 
that exhibits a discontinuous volume
change, and hence is usually considered as a first-order transition,
have puzzled the mixed-valence community for decades. They are usually
referred to as ``precursor effects'' and often associated with
gradually changing electronic properties (like ``valence''). 
Sometimes the transition is interpreted as a 
second-order one due to the softening and in contradiction to the 
obvious discontinuous behavior\cite{bkin94}. 
Our results provide a simple explanation for these precursor
effects without invoking peculiar temperature dependence of electronic
properties and clarify the rather confusing issue of the order of the
transition. This unusual transition shares 
characteristics with a second-order transition (diverging compressibility)
and a first-order transition (discontinuous order parameter).

Our analysis has been simplistic, particularly in the fact that we 
considered the two phases to be elastically isotropic and with the 
same value of the elastic coefficients, and the transition to be
coherent, namely no structural defects (dislocations) were supposed
to appear during the transition. Real materials are certainly 
anisotropic and show substantial asymmetry. 
Although probably none of the above assumptions is fully justified, we 
believe that the features obtained are quite robust. Indeed our findings are 
in good agreement with the behavior of real anisotropic mixed-valence   
systems\cite{gsc79,law81,ras78,bkin94}. 
In these systems,       %<--------
the unexpected thermodynamics is a consequence of the interplay between 
strong correlations, determining the double well potential\cite{all82,joh95}, 
and long-range strain effects. It has been suggested that volume
instabilities are generic for systems with electronic first order 
transitions\cite{lor01I,lor02}.
The present approach can be seen as a first step to 
understand this interplay in more complex non-isostructural transitions 
like the magnetoresistant manganites which also show volume
mismatch among different phases and
irreversibility\cite{cox98}. Interestingly a volume collapse
transition has recently been observed in the parent 
compound\cite{cha03}.

\section{Appendix}
In order to derive  Eq.~\eqref{Eq:fhde1k} it is convenient to work with
symmetrized strains. 
For $d=3$ we define 
$e_2=(\epsilon_{11}-\epsilon_{22})/\sqrt{2}$, 
$e_3=(\epsilon_{11}+\epsilon_{22}-2\epsilon_{33} )/\sqrt{6}$, $e_4=\sqrt{2} \epsilon_{23}$,
$e_5=\sqrt{2} \epsilon_{13}$ and 
$e_6=\sqrt{2} \epsilon_{12}$.
For $d=2$ we define 
$e_2=(\epsilon_{11}-\epsilon_{22})/\sqrt{2}$ and 
$e_3=\sqrt{2} \epsilon_{12}$. 

The shear part of the energy can be put as:
\begin{equation}
  \label{eq:fhk}
H_h= \frac{\mu}{L^d} \sum_{\bf k} \sum_{i=2}^{d(d+1)/2} e_i({\bf k}) e_i(-{\bf k})  .   
\end{equation}
The St. Venant compatibility condition leads to six equations 
in $d=3$ and to one equation in $d=2$. For simplicity we restrict to 
the $d=2$ case, the $d=3$ case follows in a similar way. The
compatibility condition reads in Fourier space: 
\begin{equation}
  \label{eq:stvenant}
C({\bf k})\equiv e({\bf k})\frac{ k^2}{\sqrt2} - e_2({\bf k})(k_x^2-k_y^2)-
2 e_3({\bf k}) k_x k_y=0  
\end{equation}
For ${\bf  k}=0$ the constrain is automatically satisfied 
and $H_h$ is minimized by  $e_2({\bf k}=0)=e_3({\bf k}=0)=0$.
In order to enforce the constraint for ${\bf  k} \ne0$ we follow
Ref.~\cite{kar95,ras01} and introduce the Lagrange 
multiplier\cite{kar95,ras01} $\lambda({\bf k})$. Minimizing 
$H_h+\sum_{{\bf  k} \ne0}\lambda({\bf k}) C({\bf k})$  
with respect to $e_2$ and $e_3$ and eliminating  $\lambda({\bf k})$ with
the aid of Eq.~\eqref{eq:stvenant} one obtains:
$e_2({\bf k})=e({\bf k})(k_x^2-k_y^2)/(\sqrt2 k^2)$ and $e_3({\bf k})=2 e({\bf
  k}) k_x k_y/(\sqrt2 k^2)$ which allows to eliminate the accommodation strains
from  Eq.~\eqref{eq:fhk} to obtain Eq.~\eqref{Eq:fhde1k}.

\end{document}